\documentclass[aps,twocolumn,amsmath,amssymb,preprintnumbers]{revtex4}
\usepackage{amsmath} \usepackage{amsfonts} \usepackage{amssymb}
\usepackage{bbm}
\usepackage{epsfig}
\usepackage{graphics}
\usepackage{graphicx}
\newcommand{\be}{\begin{equation}}
\newcommand{\ee}{\end{equation}}
\newcommand{\ba}{\begin{eqnarray}}
\newcommand{\ea}{\end{eqnarray}}
\newcommand{\nn}{\nonumber}
\newcommand{\kr}{\rangle}
\newcommand{\kl}{\langle}

\newcommand{\R}{{\cal R}}
\newcommand{\x}{(\bar x)}

\begin{document}

\title[ ]{Zwitters: particles between quantum and classical}

\author{C. Wetterich}
\affiliation{Institut  f\"ur Theoretische Physik\\
Universit\"at Heidelberg\\
Philosophenweg 16, D-69120 Heidelberg}

\begin{abstract}
We describe both quantum particles and classical particles in terms of a classical  statistical ensemble, characterized by a probability distribution in phase space. By use of a wave function in phase space both can be treated in the same quantum formalism. The different dynamics of quantum and classical particles resides then only from different evolution equations for the probability distribution. Quantum particles are characterized by a specific choice of observables and time evolution of the probability density. All relations for a quantum particle in a potential, including interference and tunneling, can be described in terms of the classical probability distribution. We formulate the concept of zwitters - particles for which the time evolution interpolates between quantum and classical particles. Experiments can test a small parameter which quantifies possible deviations from quantum mechanics.

\end{abstract}

\maketitle

Quantum particles can be described by classical statistics \cite{CWQP}. This may be viewed as a generalization of the embedding of quantum statistics in classical statistics \cite{CWAA} for the limit of infinitely many states or observables with continuous spectrum. On the other hand, the quantum formalism can be used for classical particles by introducing a real classical wave function in phase space \cite{CWQP}. The conceptual unification of quantum and classical particles permits a continuous interpolation between both ends. Zwitters are particles whose behavior is neither completely quantum nor completely classical. Experimental tests have to quantify how well quantum mechanics is obeyed by establishing bounds on a small parameter for possible deviations.

For zwitters the deviations from quantum mechanics are not related to the classical limit. They can occur for typical angular momenta or products of energy and time intervals $\Delta E\Delta t$ of the order $\hbar$. The classical limit of large $\Delta E\Delta t/\hbar$ is common to quantum particles, zwitters and classical particles. On the other hand, zwitters may correspond to an effective one-particle description of macroscopic objects as droplets of a liquid, as realized in interesting experimental settings \cite{Cov}. In this case the effective value of $\hbar $ can be much larger than the value in quantum mechanics. We concentrate in this note on a description in terms of the probability distribution $w(x,p)$ in phase space. This is the minimal common ground for a statistical description of both quantum and classical particles. More general settings for a unified discussion of quantum and classical particles will be discussed at the end of this note.

The description of quantum particles by classical probabilities may at first sight look surprising, since it seems to violate several theorems. We argue that both for classical statistics and quantum physics the outcome of several sequential or simultaneous measurements should be predicted by conditional correlation functions. Those can violate Bell's inequalities \cite{Bell}, in contrast to classical correlation functions \cite{BS}. Actually, classical correlations are not defined for the observables which describe position and momentum for quantum particles - we deal with incomplete statistics \cite{3}. Contradictions with the Kochen-Specker theorem \cite{KS} are avoided \cite{CWAA} by the appearance of equivalence classes of classical observables and by probabilistic observables \cite{CWAA}, \cite{PO}. In the present note the demonstration that no inconsistencies arise is simple: we explicitly construct all correlation functions for a quantum particle, including their time evolution, in terms of a classical probability distribution. Quantum particles are distinguished from classical particles by the use of different observables, and by a different time evolution of the probability density. For different connections between quantum and classical aspects see refs. \cite{SH}, \cite{QCC}.

\medskip\noindent
{\em Classical particles in quantum formalism} 

Our starting point is the probability distribution in phase space, $w(z,p)$, for a classical particle. It obeys the usual rules for classical probabilities
\be\label{1}
w(z,p)\geq 0~,~\int_{z,p}w(z,p)=1.
\ee
(We use $\int_z=\int d^3z,\int_p=\int d^3p/(2\pi\hbar)^3.$) The expectation values for arbitrary functions $F(z,p)$ of the classical position $z$ and momentum $p$ obtain as
\be\label{2}
\kl F(z,p)\kr=\int_{z,p}F(z,p)w(z,p).
\ee
For a particle in a potential $V(z)$ the classical time evolution of $w$ obeys the Liouville equation
\ba\label{3}
\frac{\partial}{\partial t}w=-\hat L w~,~
\hat L=\frac{p}{m}\frac{\partial}{\partial z}-
\frac{\partial V}{\partial z}\frac{\partial}{\partial p}.
\ea
(Scalar products between vectors are always assumed.) By modifying the evolution equation and introducing new types of position and momentum observables we will see that the probability density \eqref{1} can describe all aspects of a quantum particle.

It is useful to recast the probabilistic description of a particle into the quantum formalism, which can be used for classical and quantum particles as well as zwitters. An important concept is the real classical wave function $\psi_C(z,p)$, obeying
\be\label{4}
w=\psi^2_C.
\ee
This resembles the Hilbert space formulation of classical mechanics by Koopman \cite{KOP}. However, $\psi_C=s\sqrt{w}$ is here a real function which is computable from $w$. The sign function $s=\pm 1$ is essentially fixed by continuity properties \cite{CWQP}. The classical time evolution of $\psi_C$ follows from a type of Schr\"odinger equation equivalent to eq. \eqref{3},
\ba\label{5}
&&i\hbar\frac{\partial}{\partial t}\psi_C=H_L\psi_C,\nn\\
&&H_L=-i\hbar \hat L=-i\hbar \frac pm \frac{\partial}{\partial z}+i\hbar
\frac{\partial V}{\partial z}\frac{\partial}{\partial p}.
\ea
We can now employ the usual formalism of quantum mechanics by choosing commuting operators $X_{cl}$ and $P_{cl}$ for classical position and momentum. In the phase space basis they are represented by $z$ and $p$, and eq. \eqref{2} is expressed by the standard quantum formalism
\be\label{6}
\langle F(X_{cl},P_{cl})\rangle=\int_{z,p}
\psi_C(z,p)
F(z,p)\psi_C(z,p).
\ee
For classical particles eqs. \eqref{4}-\eqref{6} amount to a simple reformulation. 

\medskip\noindent
{\em Non-commuting position and momentum}

As a first step towards a quantum particle we employ quantum observables and operators for position and momentum 
\ba\label{7}
X_Q=z+\frac{i\hbar}{2}\frac{\partial}{\partial p}~,~
P_Q=p-\frac{i\hbar}{2}\frac{\partial}{\partial z}.
\ea
They differ from the classical operators and obey the commutation relation of quantum mechanics
\be\label{8}
[X^k_Q,P^l_Q]=i\hbar\delta^{kl},
\ee
similar to the Bopp operators in the context of Wigner functions. The expectation values of quantum observables are computed in terms of the classical probability distribution through the quantum expression
\be\label{9}
\kl F(X_Q,P_Q)\kr=\int_{z,p}\psi_C(z,p)F(X_Q,P_Q)\psi_C(z,p).
\ee
The order of operators now matters. We postulate that the outcome of measurements of $X_Q$ and $P_Q$ is related to $\kl F(X_Q,P_Q)\kr$ by the standard rule of the quantum formalism. For example, the dispersion of a position measurement is $\Delta^2_x=\kl X^2_Q\kr-\kl X_Q\kr^2$. This postulate can ultimately be derived from a ``microphysical ensemble'' where $X_Q$ and $P_Q$ appear as classical observables with definite values for every state \cite{CWQP,CWAA}. Eq. \eqref{9} finds a non-linear expression \cite{CWQP} in terms of the probability density $w(z,p)$ and its derivatives with respect to $z$ and $p$, such that no information beyond $w(z,p)$ is needed for the computation of $\kl F(X_Q,P_Q)\kr$.

One can introduce the ``quantum transform'' of the probability density $w(z,p)$ by employing the classical wave function $\psi_C$, 
\ba\label{10}
&&\bar\rho_w(z,p)=\\
&&\int_{r,r',s,s'}
\psi_C(z+\frac r2,p+s)\psi_C(z+\frac{r'}{2},p+s')\cos\frac{s'r-sr'}{\hbar}.\nn
\ea
In terms of $\bar\rho_w$ the totally symmetrized products of quantum observables obey a relation similar to eq. \eqref{2}
\be\label{11}
\langle F(X_Q,P_Q)\rangle=\int_{z,p}F(z,p)\bar\rho_w(z,p).
\ee
However, $\bar\rho_w$ can now be negative in certain regions of phase space and is therefore no longer a classical probability density. Eq. \eqref{11} can also be interpreted as the quantum rule for the symmetrized correlation functions of a quantum particle whose state can be fully characterized by a Wigner function \cite{Wig,Moyal} $\bar\rho_w(z,p)$. 

One may therefore ask if each possible Wigner function for a quantum particle can be obtained as a quantum transform of a suitable positive definite classical probability distribution in phase space $w(z,p)$. The answer is affirmative. It has been shown by explicit construction \cite{CWQP} that for arbitrary quantum states one can find one or several classical probability distributions $w(z,p)$ such that the quantum transform \eqref{10} yields $\bar\rho_w(z,p)$. This constitutes an explicit construction of a classical probability density from which the expectation values for quantum observables \eqref{7} can be computed according to eq. \eqref{9}, such that they coincide with the ones for a quantum particle \eqref{11} in a state given by the Wigner transform of the density matrix that can be associated to a classical wave function by eq. \eqref{10}. 

At this stage the time evolution of the probability distribution is still given by the Liouville equation for classical particles. It differs from the time evolution of a quantum particle, showing, for example, a different interference pattern in a double slit experiment. Nevertheless, we have now at our disposal two sets of observables for possible measurements of position and momentum: There are first the classical observables $X_{cl}$ and $P_{cl}$, with the function $w(z,p)$ determining expectation values according to eq. \eqref{2}. Second, for the non-commuting quantum observables $X_Q$ and $P_Q$ the expectation values can be computed in a similar fashion \eqref{11}, but now with the quantum transform $\bar\rho_w(z,p)$ instead of $w(z,p)$. It is not clear a priori which set of observables yields a better description of measurements of position and momentum of classical particles. Real measurements of classical particles are typically performed in the classical limit of large $\Delta E\Delta t/\hbar$ (or similar for some other relevant quantity). If $\hbar$-effects can be neglected, the quantum and classical observables for position and momentum coincide, cf. eq. \eqref{7}. Concerning their conceptual status the observables $X_Q$ and $P_Q$ can be associated with ``weak observables'' \cite{BH} for which measurement prescriptions have been discussed. Such non-commuting observable structures are fully compatible with a classical time evolution equation for the probability density. In order to have the analogy between classical and quantum particles as close as possible we will in the following use the quantum observables $X_Q$ and $P_Q$ also for classical particles. Other choices, including an interpolation between classical and quantum observables, are discussed in ref. \cite{CWQP}. 

\medskip\noindent
{\em Quantum evolution}

Our second step modifies the classical time evolution of the probability density \eqref{3} by a new fundamental non-linear evolution equation
\ba\label{12}
\partial_t w&=&-2\sqrt{w}L_W\sqrt{w},\\
L_W&=&\frac pm\partial_z+\frac{i}{\hbar}V\left(z+\frac{i\hbar}{2}\partial_p\right)
-\frac{i}{\hbar}V
\left(z-\frac{i\hbar}{2}\partial_p\right).\nn
\ea
This is an ordinary real first order and non-stochastic differential equation. For free particles or a harmonic potential $L_W$ coincides with $\hat L$, while for unharmonic  potentials higher order momentum derivatives appear. The difference between $L_W$ and $\hat L$ vanishes in the classical limit $\hbar\to 0$. For the associated evolution of the classical wave function one replaces in eq. \eqref{5} $H_L\to H_W=-i\hbar L_W,$
\be\label{12A}
i\hbar \partial_t\psi_C=H_W\psi_C~,~\partial_t\psi_C=-L_W\psi_C.
\ee
Since $H^\dagger_W=H_W$ and $L_W=L^*_W=-L^T_W$, the time evolution describes a rotation of the real unit vector $\psi_C$ and therefore preserves the positivity and normalization of $w$ by virtue of eq. \eqref{4}. 

Using the definition \eqref{10} one can infer from eq. \eqref{12A} that $\bar\rho_w$ obeys the same time evolution as $\psi_C$,
\be\label{13A}
i\hbar\partial_t\bar\rho_w=H_W\bar\rho_w.
\ee
This is the standard time evolution of the Wigner function for a quantum particle in a potential $V$. All predictions for expectation values \eqref{11} and their time evolution are therefore identical to quantum mechanics. This establishes that quantum particles in an arbitrary potential can be described in terms of a classical probability distribution in phase space. For a suitable choice of $w$ at some initial time $t_0$ the time evolution is such that the quantum interference pattern in a double slit experiment arises if the measurements of the location of the particle correspond to the quantum observable $X_Q$. The probability of finding the particle at the quantum position $z$ is given by $\int_p\bar\rho_w(z,p)$, while for the classical position $\bar\rho_w$ is replaced by $w$. We conclude that for a suitable choice of evolution equation \eqref{12} and observables \eqref{7} a ``classical'' positive definite probability distribution in phase space can describe all aspects of the dynamics of a quantum particle. 

In particular, pure quantum states are accounted for by classical probability distributions obeying the factorization property
\begin{eqnarray}\label{13}
&&w(z,p)=\int_{r,r'}
e^{ip(r'-r)/\hbar}\\
&&\psi^*_Q\left(z+\frac{r'}{2}\right)
\psi_Q\left(z-\frac{r'}{2}\right)
\psi^*_Q\left(z-\frac r2\right)
\psi_Q\left(z+\frac r2\right).\nn
\end{eqnarray}
Here $\psi_Q(x)$ is the usual complex Schr\"odinger wave function for the quantum particle, obeying the Schr\"odinger equation
\be\label{14}
i\hbar\frac{\partial}{\partial t}\psi_Q(x)=H_Q\psi_Q(x)~,~H_Q=-\frac{\hbar^2}{2m}
\Delta+V(x).
\ee
For pure quantum states the classical wave function equals the Wigner-transform of the quantum density matrix $\rho_Q(x,x')=\psi_Q(x)\psi^*_Q(x')$, i.e. $\psi_C(z,p)=\bar\rho_w(z,p)$. Eq. \eqref{13} amounts to an explicit construction of the classical probability distribution in phase space that is associated to a pure quantum state. 

\medskip\noindent
{\em Coarse graining}

The quantum particle can be understood as a ``coarse graining'' of the classical probability distribution. We may change the basis for the classical wave function by a Fourier transform with respect to $p$,
\be\label{15}
\tilde\psi_C(x,y)=\int_p e^{ip(x-y)/\hbar}\psi_C
\left(\frac{x+y}{2},p\right),
\ee
and introduce the ``classical density matrix''
\be\label{16}
\rho_C(x,x',y,y')=\tilde\psi_C(x,y)\tilde\psi_C^*(x',y').
\ee
We observe that the Fourier transforms $\tilde\psi_C$ and $\rho_C$ are non longer real. The classical density matrix is defined for arbitrary classical wave functions $\psi_C(z,p)$ or probability distributions $w(z,p)$.

The quantum density matrix obtains by ``integrating out'' the $y$-coordinate or ``performing a subtrace''
\ba\label{17}
\rho_Q(x,x')=\int_y\rho_C(x,x',y,y).
\ea
One may verify \cite{CWQP} that $\rho_Q(x,x')$ obeys all the formal criteria for a density matrix in quantum  mechanics. The Wigner transform of the ``coarse grained density matrix'' $\rho_Q(x,x')$, with $z=(x+x')/2$,
\ba\label{18}
\bar\rho_w(z,p)=\int d^3(x-x')e^{-ip(x-x')/\hbar}\rho_Q(x,x'),
\ea
equals the quantum transform in eq. \eqref{10}. The coarse grained density matrix $\rho_Q(x,x')$ obeys all laws for the quantum density matrix if we choose the time evolution \eqref{12A}. Insertion of this time evolution in the definition \eqref{16}, \eqref{17} yields 
\be\label{19A}
i\hbar\partial_t \rho_Q=[H_Q,\rho_Q],
\ee
with $H_Q$ given by eq. \eqref{14}. This is the quantum time evolution of the density matrix associated to the Schr\"odinger equation. Eq. \eqref{13A} for the Wigner transform can be obtained from eq. \eqref{19A} by a Fourier transform. 

The density matrix for a pure quantum state obeys the usual condition $\rho^2_Q=\rho_Q$. For this particular case one finds $\bar\rho_w(z,p)=\psi_C(z,p)$. (This relation does not hold for mixed states.) For pure states one can compute the quantum wave function $\psi_Q(x)$ from the quantum density matrix $\rho_Q(x,x')$ and therefore from $\bar\rho_w(z,p)$. In consequence, both the modulus and the phase of the complex quantum wave function can be computed from the real classical wave function and thus from the probability distribution. 

All the information necessary for the computation of expectation values of quantum observables and their quantum correlations is still available on the coarse grained level. (In the $(x,y)$-representation one has $X_{Q}=x,P_{Q}=-i\hbar \partial_x$.) In contrast, information necessary for the expectation values of classical observables and classical correlations may be lost by the coarse graining - the expectation values \eqref{6} cannot be expressed in terms of $\rho_Q(x,x')$ or $\bar\rho_w(z,p)$ alone. An exception are pure quantum states for which the expectation values and correlations for classical observables are uniquely determined by the quantum density matrix \cite{CWQP}.

\medskip\noindent
{\em Zwitters}

The concept of zwitters arises from the possibility to interpolate between the quantum and classical Hamiltonians $H_W$ and $H_L$. One of the many possible interpolations considers a time evolution of $\psi_C$ with zwitter-Hamiltonian $H_\gamma$
\be\label{20}
H_\gamma=\cos^2\gamma H_W+\sin^2\gamma H_L.
\ee
The quantum particle obtains for $\gamma=0$, while the classical particle is described by the other limit $\gamma=\pi/2$. Zwitters obtain for intermediate values of $\gamma$. The consistent definition of zwitters constitutes perhaps the most striking evidence that there is no conceptual jump between quantum and classical particles. We have no particular motivation for the specific interpolation \eqref{20} - zwitters can be defined in a much wider context by Hamiltonians that neither equal $H_L$ nor $H_W$. We stick to the specific form \eqref{20} only in order to discuss a concrete example. 

In a wider perspective, zwitters could be realized by an effective one-particle description of macroscopic bodies or collective states. For example, the dynamics of droplets of a liquid may be described in some approximation by the time evolution of a probability density in phase space that does not correspond to a classical point particle. Recent experiments \cite{Cov} point in this direction. Other interesting candidates are collective quantum states as a Bose-Einstein condensate. We will turn back to this issue at the end of this note. 

The description of a particle is consistent for arbitrary values of $\gamma$. It becomes therefore an experimental issue to quantify how well quantum mechanics is obeyed by putting limits on $\gamma$. For example, nonzero $\gamma$ will modify atomic spectra and the interference pattern in a double slit experiment. For single well isolated atoms we expect strong bounds on $\gamma$. The situation is less obvious if a large number of atoms is described by a single wave function for a ``collective particle'', as for the case of a Bose-Einstein condensate. In this case it is conceivable that nonzero $\gamma$ can be found for an effective description. We do not expect $\gamma$ to be a universal number. This parameter may rather depend on the given setting for which the notion of an isolated particle is realized. Nevertheless, the best experimental bounds on $\gamma$ can also be interpreted as fundamental tests of quantum mechanics by limiting possible deviations in a quantitative way.

Even if the world is described by perfect quantum mechanics, a lack of complete isolation or coherence may be accounted for by nonzero $\gamma$. The unitary time evolution of quantum mechanics is guaranteed only for $\gamma=0$. In contrast, the evolution of the classical wave function remains unitary for arbitrary $\gamma$. For $\gamma\neq 0$ the coarse graining  may violate the unitarity of the time evolution of the coarse grained density matrix \eqref{17} and account for phenomena as decoherence \cite{DC}, or the opposite syncoherence \cite{CWAA}, and an increasing or decreasing effective entropy for the coarse grained subsystem.

If a fundamental many body theory obeys exact quantum mechanics, the issue of a possible observation of zwitters concerns the amount of information that is necessary for the description of an ``isolated one particle state''. The notion of an isolated one particle state means that no information beyond the probability distribution $w(z,p)$ or the associated classical wave function $\psi_C(z,p)$ is available and needed for a complete description of the state. (There are easy generalizations for particles with internal degrees of freedom.) For quantum particles this information can be reduced to the information contained in the quantum density matrix $\rho_Q(x,x')$. The generalized Hamiltonian $H_W$ allows for a consistent evolution equation which does not use information beyond $\rho_Q$. If additional information, contained in $\psi_C(z,p)$ but not in $\rho_Q(x,x')$, is available and relevant for the time evolution of the isolated particle, more general evolution equations as the one for zwitters become possible. In this perspective the issue of zwitters concerns the question what type of isolated one-particle states can be realized in a quantum many body theory. The time evolution of zwitters is unitary if the information in the classical wave function $\psi_C(z,p)$ is available, but no longer unitary with respect to the coarse grained information contained in $\rho_Q(x,x')$. In this respect zwitters are distinguished from more general effects of imperfect isolation. While the isolation is not perfect in the quantum sense that information is needed beyond $\rho_Q(x,x')$, it is still realized in the classical sense that a single particle probability distribution $w(z,p)$ is sufficient for the isolated subsystem. 

We next address characteristic features of zwitters that can be tested by observation. (Cf. the second ref. \cite{CWQP} for more details.) For a quantum particle one has $[H_Q,H_W]=0$ such that $H_Q=P^2_Q/2m+V(X_Q)$ plays the role of a conserved energy, similar to the classical energy $H_{cl}=P^2_{cl}/2m+V(X_{cl})$ which commutes with $H_L$. No such conserved energy is available for zwitters for $\gamma\neq 0,\pi/2$. (The generator of time translations is $H_\gamma$ which vanishes for all static states.) This observation may serve for establishing experimental bounds on $\gamma$. Indeed, a typical ground state for a zwitter has no sharp energy but rather a nonzero width. In turn, eigenstates of the quantum  energy $H_Q$ are not static since $[H_Q,H_\gamma]\neq 0$ (except for harmonic $V$). Let us assume from now on that position, momentum and energy for zwitters are measured by the quantum observables $X_Q,P_Q$ and $H_Q$, such that the only difference to a quantum particle arises through the modified time evolution for $\gamma\neq 0$. 

A good candidate for the ground state of a zwitter is a probability distribution which leads to a static coarse grained density matrix $\rho_Q(x,x')$. Among these ``coarse grained static states'' we consider the one with lowest $\langle H_Q\rangle$. For small $\gamma$ it is approximately given by a pure quantum state $\psi^{(\gamma)}_0(x)$. This wave function is dominated by the quantum ground state $\psi_0$, but has small admixtures $\sim \sin^2\gamma$ of higher energy eigenstates $\psi_n (H_Q\psi_n=E_n\psi_n$). We find a nonzero energy width for the zwitter ground state
\be\label{20A}
\Delta E=(\langle H^2_Q\rangle-\langle H_Q\rangle^2)^{1/2}=f_1\sin^2\gamma|E_0|,
\ee
where the calculable constant $f_1$ depends on the potential. (For a Coulomb potential one has $2<f_1<10$.) Also the mean energy of the ground state increases for $\gamma>0$
\be\label{20B}
\langle H_Q\rangle=E_0+\delta E^{(\gamma)}~,~\delta E^{(\gamma)}=f_2\sin^2\gamma\Delta E.
\ee
However, this shift $(0.5\leq f_2\leq 2.5$ for a Coulomb potential) is much smaller than the width $\Delta E$. The nonzero width $\Delta E$ of the ground state contrasts with the quantum particle. Measurements of $\Delta E$ may therefore determine $\gamma$ or yield bounds. As an example, one may extract $\Delta E<0.6\cdot 10^{-20}$ eV from  the relaxation time of nuclear polarized $^3$He \cite{Gem}, and estimate by comparison with the binding energy $|\gamma|\lesssim 3\cdot 10^{-14}$. (A more detailed calculation of the system with a zwitter Hamiltonian would be necessary in order to extract a precise bound.) A particularly interesting situation arises if the first excited quantum state is very close to the quantum ground state such that $\Delta E$ and $E_1-E_0$ are of comparable order. 

For small $\gamma$ we may use in eq. \eqref{20} $\cos^2\gamma=1$, $\sin^2\gamma=\gamma^2=\epsilon$. More general zwitter Hamiltonians for small deviations from quantum mechanics can be obtained by
\be\label{22A}
H_\epsilon=H_W+\epsilon H_\Delta,
\ee
with $H_\Delta$ involving $z,p$ as well as derivatives $\partial_z,\partial_p$. For a consistent time evolution of the classical wave function $H_\Delta$ must be hermitean and purely imaginary, such that $L_\Delta=(i/\hbar)H_\Delta$ adds to the antisymmetric generator of rotations, $L_W+\epsilon L_\Delta$. 

\medskip\noindent
{\em Conceptual issues of quantum mechanics}

Having derived all features of standard quantum mechanics (for $\gamma=0$) from a classical probability distribution in phase space, a few thoughts about conceptual issues may be in order. Since all correlation functions for the quantum observables $X_Q$ and $P_Q$ follow the quantum rules, there is no doubt that measurements should yield the values predicted by quantum mechanics if we use the same correspondence between the distribution of measurement values and correlation functions. In a sequence of two measurements for observables $A$ and $B$ the appropriate conditional correlation function $\kl BA\kr_m$ multiplies first the possible measurement value $A_\alpha$ with the probability $w_\alpha$ for the states for which the observable $A$ has the value $A_\alpha$. This is then multiplied with the possible measurement value $B_\beta$ and the {\em conditional} probability $(w_\beta|\alpha)$ to find a value $B_\beta$ {\em if} $A_\alpha$ has been found in the first measurement. One finally sums over all $\alpha$ and $\beta$
\be\label{Z1}
\kl BA\kr_m=\sum_{\alpha,\beta}A_\alpha B_\beta(w_\beta|\alpha)w_\alpha.
\ee
We have advocated that this correlation function is given by the anticommutator of the associated quantum operators $\hat A,\hat B$ \cite{CWAA},
\be\label{Z2}
\kl BA\kr_m=\frac12\kl\{\hat A,\hat B\}\kr.
\ee
While two measurements commute, the order matters for a generalization to three measurements \cite{CWAA}. The conditional correlations \eqref{Z1} can violate Bell's inequalities \cite{CWAA}. In contrast, Bell's inequalities would follow if we would replace in eq. \eqref{Z1} the conditional probability $(w_\beta|\alpha)w_\alpha$ by the joint probability $w_{\beta\alpha}$ of finding $B_\beta$ and $A_\alpha$ in a given state of the ensemble. Joint probabilities for $X_Q$ and $P_Q$ are not defined in terms of $w$ or $\bar\rho_w$, however.

We emphasize that we have obtained quantum mechanics from the concept of a classical statistical ensemble, but not from a deterministic theory based on trajectories. Indeed, Newtonian trajectories would lead to the Liouville equation and not be compatible with the time evolution \eqref{12}. Our basic setting is probabilistic realism. There is one reality, but only a probabilistic description is possible and meaningful. Physics describes conditional probabilities for sequences of events. In the presence of correlations a system cannot be divided into independent subsystems - the whole is more than the sum of its parts. This also holds for non-local correlations between regions that cannot exchange light signals. Non-local correlations are common in statistical physics, a good example being the anisotropies in the cosmic microwave background. A causal theory only requires that non-local correlations have been generated by causal events in the past. No paradoxon arises in the Einstein-Rosen-Podolski setting \cite{EPR} if one limits the discussion to these probabilistic concepts \cite{CWAA}. There is no contradiction to realism and causality if correlations are accepted as genuine part of reality. 

A further conceptual point may become apparent if we express the expectation value of the squared quantum momentum in terms of the probability density
\be\label{21}
\langle P^2_Q\rangle=\langle P^2_{cl}\rangle+\frac{\hbar^2}{16}\langle (\partial_z\ln w)^2\rangle.
\ee
The second contribution is of statistical nature, similar to quantities like the entropy in an equilibrium ensemble. It involves a phase space integral over $(\partial_zw)^2/w$ and is therefore not linear in $w$, as opposed to the expectation values of classical observables. In a sense, the second part measures the ``roughness'' of the distribution in position space. Together with a similar contribution in $\kl X^2_Q\kr$ for the roughness in momentum  space, this is responsible for Heisenberg's uncertainty relation.

While an understanding of several important conceptual issues of quantum mechanics is possible {\em within} our description in terms of a probability density $w(z,p)$ in phase space, this does not hold for all of them. In particular, particle-wave duality is not addressed in this limited setting. In quantum mechanics ``interval observables'' are defined by functions
\be\label{A1}
J_\R(\bar x)=\left\{\begin{array}{ll}
1&\text{for }x\in \R\\
0&\text{otherwise}
\end{array}\right.,
\ee
with $\R$ some region of space (interval) around $\bar x$. The expectation value is simply given by
\be\label{A2}
\kl J_\R(\bar x)\kr=\int_V\psi^*(x)J_\R(\bar x)\psi(x)=\int_\R\psi^*(x)\psi(x),
\ee
where the integrals $\int_V$ and $\int_\R$ are space integrals over the whole volume $V$ or the region $\R$, respectively. Since $J_\R\x$ is a projector, $J^2_\R\x=J_\R\x$, its spectrum consists of the values zero and one. According to the rules of quantum mechanics the possible measurement values of such interval observables are only zero or one. This has a simple interpretation: either the particle is present in the region around $\bar x$, yielding $J_\R\x=1$, or it is somewhere else, with $J_\R\x=0$. This discreteness of particles, together with a continuous probability distribution describing how likely it is to find the particle in the region around $\bar x$, constitutes the essence of particle-wave duality. 

No sign of the discrete particle properties is visible in the description by $w(z,p)$. We can implement the interval observables $J_\R\x$ in our classical statistical setting. Their expectation values will indeed obey the quantum formula \eqref{A2}. However, in order to account for the restricted spectrum of possible measurement values zero or one we have to associate $J_\R\x$ with {\em probabilistic observables} as discussed in ref. \cite{CWAA}. Typically, probabilistic observables arise from a classical statistical ensemble with a larger number of ``substates'', for which part of the information is already integrated out on the level of the one-particle probability distribution $w(z,p)$.

In fact, a full description of quantum particles by classical statistics involves two crucial aspects. First, one needs to obtain the dynamics of the Schr\"odinger equation from an evolution equation of a classical probability density or associated classical wave function. This is achieved in the present note. Second, one needs a formulation in terms of classical states for which the observables can only take the values allowed by quantum mechanics. In the present note this poses no problem for the continuous observables position and momentum, since they can assume arbitrary real values. For a realization of the interval observables $J_\R\x$ one needs, however, to consider some ``underlying level'', as sketched, in principle, in the first ref. \cite{CWQP}.

Obtaining the Schr\"odinger equation in a classical statistical setting (without implementing the discrete particle aspects) can be achieved in a wide variety of settings. As a simple example, one can derive the Schr\"odinger equation as a classical field equation from the Klein-Gordon equation in an external electromagnetic field,
\be\label{A3}
(\eta^{\mu\nu}D_\mu D_\nu-m^2)\varphi=0,
\ee
where $\partial_0=\partial_t$ (we use $c=1), ~\eta^{\mu\nu}=diag(-1,1,1,1)$, and 
\be\label{A4}
\varphi=
\left(\begin{array}{c}\varphi_1\\ \varphi_2\end{array}\right)~,~
D_\mu\varphi=(\partial_\mu+eA_\mu I)\varphi~,~I=
\left(\begin{array}{ccc}0&,&-1\\1&,&0\end{array}\right).
\ee
For a simple demonstration that the phases of quantum mechanics arise from the presence of a complex structure in a real formulation we employ here two real fields $\varphi_{1,2}$. The usual complex Klein-Gordon equation obtains for the complex scalar field $\varphi=\varphi_1+i\varphi_2$. The form of the Klein-Gordon equation is determined by Lorentz-symmetry and the number of derivatives. It is a classical field equation with the same status as Maxwell's equations for electromagnetism. As is well known, the non-relativistic approximation yields the Schr\"odinger equation for a particle with mass $M=\hbar m$ in a potential $V(x)=e\hbar A_0(x)$. The Schr\"odinger wave function $\psi(x)$ is defined as $\psi(x)=\exp (-imt)\varphi(x)$. One observes the appearance of $\hbar$ as pure conversion factor for units of mass and charge. The interpretation of quantum mechanics as a classical system in ref. \cite{SH} corresponds to this picture of the Schr\"odinger equation as a classical field equation. 

One could build a classical statistical description on the classical field variables $\tilde\varphi_i(x)$, with probability distribution $w\big[\tilde\varphi_1(\vec x),\tilde\varphi_2(\vec x)\big]$. The limiting case of a deterministic classical field theory is realized by a ``sharp probability distribution'' $w\big(\big[\tilde\varphi_1(\vec x),\tilde\varphi_2(\vec x)\big],t\big)$. For any time $t$ this differs from zero only for one precise field value $\tilde\varphi_{1,2}(\vec x)=\varphi_{1,2}(\vec x,t)$, where the ``classical field'' $\varphi_{1,2}(\vec x,t)$ obeys the Klein-Gordon equation. However, the field product $\psi^*(x)\psi(x)$ is now given by a continuous classical observable. Thus the r.h.s. of the expression for the interval observable \eqref{A2} can take continuous values, not only zero or one as required by quantum mechanics. This holds both for a ``deterministic setting'' of a classical field equation and for the probabilistic generalization. If the fields $\psi(x)$ are associated with the classical states, the interval observables $J_\R(x)$ are classical observables in the usual sense that they have a fixed value for every classical state. Since the spectrum of these observables is continuous, it seems hard to explain why the observed values in measurements of such observables should only yield zero or one. Quantum mechanics differs in this important aspect from a classical field theory.

\medskip\noindent
{\em Zwitters from quantum field theory?}

Recently, a quantum field theory for Dirac fermions in an arbitrary external electromagnetic field has been obtained from a suitable evolution equation for a classical statistical ensemble of Ising-spins \cite{CWQFT}. For every point on a space-lattice the Ising spins can only take two values, corresponding to occupation numbers or bits with values zero or one. The discrete occupation numbers of a multi-fermion system can also take only the values zero and one. They are connected directly with the Ising spins. For this classical statistical ensemble the interval observables $J_\R\x$ can be realized as classical observables \cite{CWQFT}. The discrete values of $J_\R\x$ are now directly related to the discrete values of the Ising spins. The configurations of Ising spins are therefore a good candidate for the ``substates'' that could give rise to probabilistic observables on the level of a reduced probability distribution for isolated particles. 

The quantum particle in a potential has been realized in this classical statistical Ising-type model. A conserved particle number allows the definition of one-particle states. The associated one-particle wave function obeys the Dirac equation in an external electromagnetic field. The non-relativistic approximation yields the Schr\"odinger equation for a particle in a potential \cite{CWQFT}. This classical statistical ensemble therefore realizes particle-wave duality. The discrete spectrum of particle observables as $J_\R\x$ is linked to the discreteness of the Ising-spins, and the continuous wave properties refer to the continuous probability distribution for the discrete configurations or classical states. (In this approach $\psi_Q(x)$ is not a classical field that characterizes the states of the classical statistical ensemble or a deterministic classical field theory.) The classical statistical ensemble of Ising-spins fully accounts for a quantum particle - both for the dynamics of the Schr\"odinger equation and the discrete particle properties. 

One may ask if zwitters can also be realized in such a setting. Of course, one could modify the fundamental evolution equation for the probability distribution of the Ising-type model. This would yield deviations from quantum mechanics. Perhaps even more interesting is the possibility of obtaining zwitters within the framework of a quantum many-body system. In this case one maintains the evolution equation proposed in ref. \cite{CWQFT} such that all quantum properties for a system of Dirac fermions in electromagnetic fields are realized. Within such a setting one may now focus on possible isolated one-particle states. This amounts to a coarse graining of the information. Instead of the the full probability distribution for all possible configurations of Ising spins, which are equivalent to arbitrary states of an arbitrary number of fermions, one concentrates on a reduced statistical ensemble where the states only reflect one-particle properties. The reduced probability distribution associates a probability to every state of the reduced ensemble. Isolation means that the time evolution of the reduced wave function can be expressed in terms of the reduced wave function alone, without invoking other details of the wave function for the full system. (The time evolution of an isolated system should be independent of the ``environment''.)

A minimal setting of a reduced ensemble for a single particle involves the position $x$ and a discrete internal property that can take two values, $\gamma=1,2$. The real classical wave function $\psi(x,\gamma)$ contains sufficient information for the construction of a complex wave function, $\psi_Q(x)=\psi(x,1)+i\psi(x,2)$. Such a construction of an isolated one-particle state has been given explicitly in ref. \cite{CWQFT} and describes a quantum particle. For the realization of an isolated zwitter particle or classical particle an extended variety of states for the reduced ensemble is necessary. The reduced states are now characterized by $(z,p)$ or $(x,y)$, i.e. two position variables for one isolated particle. As compared to the quantum particle this indicates some internal structure, since the classical wave function $\tilde\psi_C(x,y)$ depends not only on the ``center of mass position'' $z$, but also on some relative coordinate $x-y$. (The relation $\tilde \psi^*_C(x,y)=\tilde\psi_C(y,x)$ guarantees that $\tilde\psi_C(x,y)$ contains not more information than a real function of six variables $(\vec x,\vec y)$.) At the present stage is not known if isolated subsystems of this type exist for the classical statistical ensemble of Ising-spins. It is remarkable that in the Ising-type classical statistical model it seems much easier to realize an isolated quantum particle than an isolated classical particle. 

\medskip\noindent
{\em Zwitters as collective ``one-particle states''}

The formalism employed in this note could find applications for a wide variety of classical or quantum collective systems. So far we have concentrated on quantum particles or zwitters close to quantum particles. The issue that the time evolution of an effective one-particle system may correspond neither to a quantum nor to a classical particle is much more general, however. Let us assume a situation where the relevant states of a collective many-body system can be characterized by a position type variable $z$ and a momentum type variable $p$. An example are the center of mass coordinate and the momentum of a water droplet. (We do not discuss here possible generalizations to effective one-particle states with internal degrees of freedom.) We choose a probabilistic description where the probability distribution $w(z,p)$ or the associated classical wave function $\psi_C(z,p)$ obtains by integrating over all other properties of the collective system except $z$ and $p$. In other words, $w(z,p)$ is computed by summing the probabilities of all microscopic states of the collective system for which the position and momentum observables have common values $z$ and $p$. (In certain instances the coarse graining of information within a classical statistical ensemble yields a classical density matrix $\rho_C(z,p,z',p')$ that does not correspond to a pure state with classical wave function $\psi_C(z,p)$ \cite{CWAA}. Generalization to this case is possible.)

An effective ``one-particle state'' is realized if the time evolution of $\psi_C(z,p)$ can be written as
\be\label{33}
\partial_t\psi_C(t)=F\big[\psi_C(t)\big].
\ee
Here we concentrate on a causal evolution where the functional $F\big[\psi_C(t)\big]$ depends only on the wave function at time $t$. An evolution of this type \eqref{33} means that only information about probabilities of the reduced system in the form of $\psi_C(z,p;t)$ is needed for determining $\psi_C$ at a following time $t+dt$. (This may often be only a reasonable approximation.) The special case of a linear evolution
\be\label{34}
\partial_t\psi_C(t)=-L\psi_C(t)
\ee
covers both the classical and the quantum particle. We have concentrated in this note on this simple case. Any $L$ different from the Liouville operator $\hat L$ or from $L_W$ realizes a zwitter. (The more general case \eqref{33} may be called a non-linear zwitter.) 

The crucial part in the evolution of such collective systems will be the response to an external potential - only this distinguishes a quantum and a classical particle in eqs. \eqref{12A} and \eqref{5}. It remains to be seen if an effective evolution generator $L_W$ in eq. \eqref{12} can be realized in a suitable macroscopic experimental setting. (The value of $\hbar$ in eq. \eqref{12} would be replaced in this case by an effective macroscopic constant.) The remarkable behavior of droplets in the experiments reported in ref. \cite{Cov} suggests objects similar to zwitters. 

In conclusion, we have demonstrated that not only classical particles but also quantum particles can be described by a classical statistical ensemble with a suitable time evolution law. In this note we have concentrated on isolated one-particle states that are described by a probability distribution or classical wave function in phase space. In this setting the common formalism for classical and quantum particles allows for a continuous interpolation by zwitters. On the one hand, a consistent formulation of a probabilistic theory that is arbitrarily close to quantum mechanics allows for quantitative experimental tests of the validity of quantum theory by establishing bounds on a small parameter, rather than yes/no tests between classical and quantum. On the other hand, our findings raise the interesting question if isolated ``one-particle states'' with the time evolution of zwitters could be realized within a classical or quantum many-body theory. Such states could be collective states as droplets of a liquid or a Bose-Einstein condensate. 

\medskip\noindent
Acknowledgment: The author thanks D. Dubbers for communicating ref. \cite{Gem}.

\end{document}